\documentclass[pre,twocolumn]{revtex4-2}
\usepackage{graphicx} 

\usepackage{amsmath,amsthm,amssymb,mathrsfs}
\usepackage{bm}

\begin{document}

\title{Step-like dependence of memory function on pulse width in spintronics reservoir computing}

\author{
  Terufumi Yamaguchi${}^{1}$, Nozomi Akashi${}^{2}$, Kohei Nakajima${}^{2}$, Hitoshi Kubota${}^{1}$, Sumito Tsunegi${}^{1}$, and Tomohiro Taniguchi${}^{1}$
      }
 \affiliation{
 ${}^{1}$National Institute of Advanced Industrial Science and Technology (AIST), Spintronics Research Center, Tsukuba, Ibaraki 305-8568, Japan, \\
 ${}^{2}$Graduate School of Information Science and Technology, The University of Tokyo, Bunkyo-ku, 113-8656 Tokyo, Japan
 }

\date{\today} 
\begin{abstract}
{
Physical reservoir computing is a type of recurrent neural network that applies the dynamical response from physical systems to information processing. 
However, the relation between computation performance and physical parameters/phenomena still remains unclear. 
This study reports our progress regarding the role of current-dependent magnetic damping in the computational performance of reservoir computing. 
The current-dependent relaxation dynamics of a magnetic vortex core results in an asymmetric memory function with respect to binary inputs. 
A fast relaxation caused by a large input leads to a fast fading of the input memory, 
whereas a slow relaxation by a small input enables the reservoir to keep the input memory for a relatively long time.
As a result, a step-like dependence is found for the short-term memory and parity-check capacities on the pulse width of input data, 
where the capacities remain at 1.5 for a certain range of the pulse width, and drop to 1.0 for a long pulse-width limit. 
Both analytical and numerical analyses clarify that the step-like behavior can be attributed to the current-dependent relaxation time of the vortex core to a limit-cycle state. }
\end{abstract}

 \maketitle



A recurrent neural network (RNN) is categorised as a class of artificial neural networks having recurrent interactions between a large number of neurons \cite{mandic01}. 
The RNN can memorise data inputted into the system through recurrent interactions. 
The dynamical response from the network therefore reflects a time sequence of the input data. 
In this respect, RNNs have attracted much attention not only in fundamental sciences 
but also applied sciences such as information processing of time-dependent data, for example, human voices and robotic motions. 
Physical reservoir computing is a model of RNNs 
where many-body systems, called reservoirs, are used as the networks \cite{maas02,jaeger04,verstraeten07,hermans10,appeltant11,grigoryeva18,rohm18,nakajima20}, 
enabling it to bridge the gap between neural science, information science, biology, and physics. 
Physical reservoir computing have been performed in several kinds of physical systems, such as optical lasers, soft materials and quantum reservoirs 
\cite{brunner13,nakajima15,fujii17,dion18,nakajima19}. 


Recent studies have discovered that a fine-structured ferromagnet can also be applied to physical reservoir computing 
\cite{torrejon17,furuta18,tsunegi18,bourianoff18,nakane18,nomura19,markovic19,tsunegi19,riou19,yamaguchi20}. 
By applying an electric current to a magnetic multilayer in nanoscale, spin transfer \cite{slonczewski96,berger96} from conducting electrons to a local magnetisation 
induces nonlinear magnetisation dynamics such as magnetisation switching and limit-cycle oscillation \cite{katine00,kiselev03,rippard04,kubota13}. 
Such a dynamical response from the ferromagnet is divided into several nodes, with each node regarded as a virtual neuron. 
The method to construct a virtual many-body system, in this case, a reservoir, is called a time-multiplexing method \cite{fujii17,torrejon17,tsunegi18}. 
The high performance of a voice recognition using a vortex-type spin-torque oscillator \cite{torrejon17} has evidently proven that spintronic systems are effective for physical reservoir computing. 
While such exciting works for practical applications have been reported, the fundamental properties of physical reservoir computing have yet to be fully clarified. 
For example, the role of the physical parameters, such as the current magnitude and magnetic damping constant, on the performance of physical reservoir computing remains unclear. 
Further development of physical reservoir computing, or a wide range of brain-inspired computing in general, 
relies crucially on the clarification of the relation between the physical parameters and computation performances.


In this work, a theoretical study is carried out for physical reservoir computing using a vortex-type ferromagnet. 
The short-term memory (STM) and parity-check (PC) capacities are evaluated as a function of the pulse width of input data given by current pulses. 
The capacities show a step-like dependence on the pulse width, 
where the capacities' values remain at $1.5$ for a certain range of the pulse width, and drop to $1.0$ for a long pulse-width limit, 
where the range of the pulse width depends on the material parameters and input-current strength. 
It is clarified that the step-like behaviour originates from the current-dependent relaxation time of the vortex core. 


\section*{Model}



\begin{figure}
\centerline{\includegraphics[width=1.0\columnwidth]{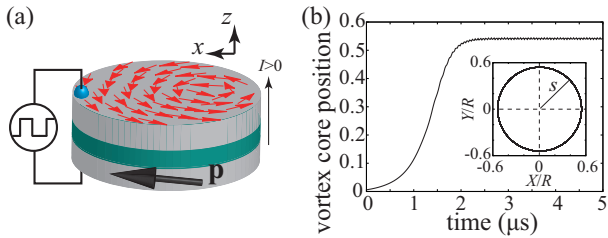}}
\caption{
         (a) Schematic diagram of the magnetic multilayer consisting of a vortex-type free layer and a uniformly-magnetised reference layer. 
         (b) Time evolution of the normalised centre position of the vortex core in the presence of a direct current. 
             The inset shows a limit-cycle oscillation of the vortex core in a steady state. 
         \vspace{-3ex}}
\label{fig:fig1}
\end{figure}



Figure \ref{fig:fig1}(a) schematically shows a magnetic trilayer consisting of a vortex-type free layer and a uniformly-magnetised reference layer separated by a thin nonmagnetic spacer. 
The $z$ axis is normal to the film-plane. 
The unit vector pointing in the magnetisation direction of the reference layer is denoted as $\mathbf{p}=(p_{x},0,p_{z})$ ($|\mathbf{p}|=1$), 
where the $x$ axis is chosen to be parallel to the projection of $\mathbf{p}$ to the $xy$ plane. 
The dynamics of the magnetic vortex core are known to be well described by the Thiele equation \cite{thiele73,guslienko06PRL,guslienko06,khvalkovskiy09,guslienko11,dussaux12}. 
By applying an electric current density $J$, the spin-transfer torque \cite{slonczewski96,berger96} induces the dynamics of the vortex core described by 
the Thiele equation for the core position $\mathbf{X}=(X,Y,0)$ \cite{thiele73,guslienko06PRL,guslienko06,khvalkovskiy09,guslienko11,dussaux12}, 
\begin{widetext}
\begin{equation}
\begin{split}
  &
  -G \hat{\mathbf{z}}
  \times
  \dot{\mathbf{X}}
  -
  |\mathscr{D}| 
  \left(
    1
    +
    \xi s^{2}
  \right)
  \dot{\mathbf{X}}
  -
  \frac{\partial W}{\partial \mathbf{X}}
  +
  a_{J} J p_{z}
  \hat{\mathbf{z}}
  \times
  \mathbf{X}
  +
  c a_{J} J R_{0} p_{x}
  \hat{\mathbf{x}}
  =
  \bm{0},
  \label{eq:Thiele}
\end{split}
\end{equation}
\end{widetext}
where $G=2\pi pc ML/\gamma$ and $\mathscr{D}=-(2\pi \alpha ML/\gamma)[1- (1/2)\log(R_{0}/R)]$ consist of 
the saturation magnetisation $M$, the gyromagnetic ratio $\gamma$, the Gilbert damping constant $\alpha$, 
the thickness $L$, the disc radius $R$ and the core radius $R_{0}$ of the free layer. 
The polarity $p$ and the chirality $c$ are assumed to be $+1$ for convenience. 
The normalised centre position of the vortex core is $s=|\mathbf{X}|/R$. 
The nonlinear parameter of the damping torque is denoted as $\xi$. 
The magnetic potential is 
\begin{equation}
  W
  =
  \frac{\kappa}{2}
  |\mathbf{X}|^{2}
  +
  \frac{\kappa^{\prime}}{4R^{2}}
  |\mathbf{X}|^{4}, 
  \label{eq:magnetic_energy}
\end{equation}
where $\kappa=(10/9)4\pi M^{2}L/R$ and $\zeta=\kappa^{\prime}/\kappa\simeq 1/4$ \cite{guslienko06PRL,dussaux12}. 
The spin-transfer torque strength with the spin polarisation $P$ is $a_{J}=\pi\hbar P/(2e)$. 
A positive current corresponds to the current flowing from the reference layer to the free layer. 
The values of the parameters are estimated from experiments and simulations \cite{dussaux12,grimaldi14,tsunegi14,tsunegi16} as 
$M=1500$ emu/cm${}^{3}$, $\gamma=1.764\times 10^{7}$ rad/(Oe s), $\alpha=0.005$, $L=4$ nm, $R=150$ nm, $R_{0}=10$ nm, $\xi=1/4$, and $P=0.3$. 

The vortex dynamics are described by two dynamical variables, which are $(X,Y)$ in a Cartesian coordinate or 
the normalised distance $s=|\mathbf{X}|/R$ of the core centre from the disc centre and the phase of the core position in the $xy$ plane. 
In previous experiments on reservoir computing in Refs. \cite{torrejon17,tsunegi18}, however, only $s$ is used for the computing. 
Therefore, we are also inclined to focus on the dynamics of $s$ in the following discussion. 
Simultaneously, we note that recent works have focused on reservoir computing using the phase of the vortex oscillation \cite{markovic19,tsunegi19}. 

Figure \ref{fig:fig1}(b) shows the time evolution of the vortex core position $s$ in the presence of a direct current of $I=\pi R^{2}J_{\rm dc}=4.0$ mA. 
The vortex core position $s$ is saturated after a time on the order of $1$ $\mu$s, 
and a limit-cycle oscillation of the vortex core around the disc centre is excited, as shown in the inset. 
The relaxation phenomenon plays a key role on the memory function of reservoir computing, as discussed below. 
We also note that the core position in the limit-cycle state is not constant 
because the spin-transfer torque originating from the in-plane component of the magnetisation $\mathbf{p}$ breaks the rotational symmetry of the vortex core around the $z$ axis. 
As a result, the distance $s$ of the vortex core from the disc centre for the circular motion is not constant, as shown in Fig. \ref{fig:fig1}(b). 
We note that a small-amplitude oscillation disappears when $p_{x}$ in Eq. (\ref{eq:Thiele}) is zero. 
In the experiments of Refs. \cite{torrejon17,tsunegi18,markovic19,tsunegi19}, however, $p_{x}$ remains finite 
because the output signal from the oscillator, generated through the tunnel magnetoresistance effect, is determined by the number of magnetic moments having the projection to the $\mathbf{p}$ direction. 
On the other hand, the critical current density to excite the vortex motion is determined by $p_{z}$, as shown below. 
Therefore, we keep both terms proportional to $p_{x}$ and $p_{z}$ in Eq. (\ref{eq:Thiele}) finite. 
In the experiments, we apply a small external magnetic field pointing in the $z$ direction to the in-plane magnetised reference layer 
and make both $p_{x}$ and $p_{z}$ finite \cite{torrejon17,tsunegi18,markovic19,tsunegi19}.


Reservoir computing in the system used in this study is performed by applying random binary pulse inputs to the magnetic trilayer. 
For example, the input data was encoded in the amplitude of the direct current \cite{torrejon17,tsunegi18} or microwave magnetic field \cite{tsunegi19} in previous works.  
The input data in this work is encoded in the amplitude of the direct current as 
\begin{equation}
  J
  =
  J_{\rm dc}
  \left[
    1
    +
    \nu 
    {\rm bi}(t)
  \right],
  \label{eq:current}
\end{equation}
where $J_{\rm dc}$ is the magnitude of the direct current density and $\nu$ is the ratio of the binary pulse with respect to the bias current density $J_{\rm dc}$. 
In this work, we use $\nu=0.2$, except the results shown in Fig. \ref{fig:fig5} where $\nu=0.05$. 
The random binary data is ${\rm bi}(t)=0\ {\rm or}\ 1$, which is constant during a pulse width $t_{\rm p}$. 



\section*{Results}


\subsection*{Dynamical response to random input}
\label{sec:Dynamical response to random input}

\begin{figure*}
\centerline{\includegraphics[width=2.0\columnwidth]{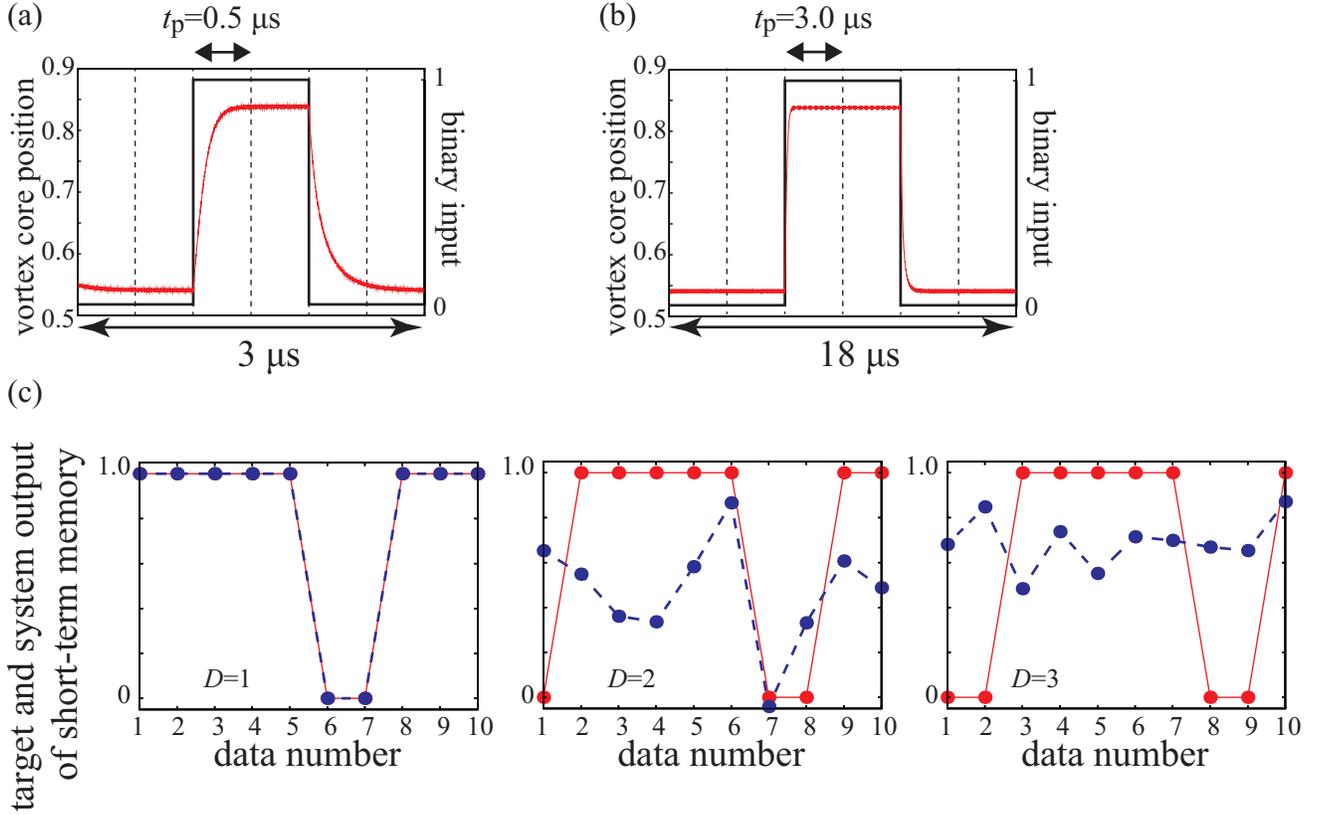}}
\caption{
         Time evolutions of the vortex-core position (red) and binary pulses (black) with pulse widths of (a) $0.5$ $\mu$s and (b) $3.0$ $\mu$s. 
         The six binary pulses denote the input "001100". 
         The dotted line represents unit pulse width $t_{\rm p}$. 
         (c) Target output (red, solid) and system output (blue, dashed) of the STM task. 
         The delay varies as $D=1,2$, and $3$ from left to right. 
         The pulse width is $3.0$ $\mu$s. 
         \vspace{-3ex}}
\label{fig:fig2}
\end{figure*}

Figures \ref{fig:fig2}(a) and \ref{fig:fig2}(b) show the dynamics of the centre position of the vortex core $s(t)$ from the random binary input 
with pulse widths of $0.5$ $\mu$s and $3.0$ $\mu$s, respectively. 
Six pulses, divided by dotted lines, represent the binary data "001100" as shown by the black line for example. 
The core position saturates to the value shown in Fig. \ref{fig:fig1}(b) when the binary pulse of ${\rm bi}=0$ is inputted, 
whereas it saturates to a large value when a binary pulse of ${\rm bi}=1$ is inputted 
because the spin-transfer torque due to the additional current, $\nu J_{\rm dc}$, pushes the vortex core away from the disc centre. 
As a result, the relaxation dynamics between the two states is observed when the value of the binary pulse changes. 
When the pulse width is shorter than the time to saturate the vortex core position, a gradual change in the vortex-core position can be observed, as shown in Fig. \ref{fig:fig2}(a). 
On the other hand, the vortex core remains almost constant during a pulse when the pulse width is longer than the time to saturate the vortex core position, as shown in Fig. \ref{fig:fig2}(b).



\subsection*{Evaluation of short-term memory and parity-check capacities}
\label{sec:Evaluation of short-term memory and parity-check capacities}

As a figure of merit for reservoir computing, we evaluate the STM and PC capacities. 
The memory capacity characterises the number of data points the reservoir can store and is a dimensionless quantity. 
The STM capacity is the memory capacity for a linear combination of the input data as defined in Eq. (\ref{eq:target_data_STM}) below. 
The PC capacity is the memory capacity for nonlinear data, where a nonlinear transformation is applied to the input data, as shown in Eq. (\ref{eq:target_data_PC}) below. 
In general, the memory capacity depends on the sequence of the input data, 
therefore, it is defined as an average of the number of stored data with respect to random input. 


The time-multiplexing method is applied to construct a virtual many-body system from the dynamical response of a single ferromagnet. 
The dynamical response $s(t)$ during a pulse is divided into $N_{\rm node}$, 
and $s_{k,i}$ corresponding to $s(t)$ at the $i$th node in the presence of the $k$th pulse is regarded as the $i$th neuron at a discrete time $k$ \cite{fujii17,torrejon17,tsunegi18}. 
By applying $N=1000$ pulses and dividing the output $s(t)$ into $N_{\rm node}=250$ nodes, 
we determined the weight $w_{D,i}$ minimising the error between the system output and the target data given by 
\begin{equation}
  \sum_{k=1}^{N}
  \left(
    \sum_{i=1}^{N_{\rm node}+1}
    w_{D,i} 
    s_{k,i}
    -
    v_{k,D}
  \right)^{2},
\end{equation}
where the bias term is $s_{k,N_{\rm node}+1}=1$ \cite{fujii17}. 
The integer $D(=0,1,2,\cdots)$ represents the delay \cite{fujii17,furuta18,tsunegi18,tsunegi19}. 
Since reservoir computing calculates the time sequence of input data, the capability of a reservoir is characterised by the number of past input data stored in it. 
The delay $D$ characterises such past input data; for example, it shows whether the $(k-D)$th input data can be reproduced from the $k$th output. 
Accordingly, the delay $D$ is a quantity related to the order of the input data, and thus, is an integer. 
It does not relate to any time scale of physical systems, such as physical delay time. 
In this paper, the symbol $D$ appeared in quantities related to reservoir computing and is used as an integer represents the delay, 
whereas the symbol $\mathscr{D}$ that appeared in the Thiele equation represents the damping strength.
The data the reservoir should reproduce is called target data or output, and is constructed from the input data. 
The target outputs for the STM and PC tasks are, respectively, given by 
\begin{equation}
  v_{k,D}^{\rm STM}
  =
  {\rm bi}_{k-D},
  \label{eq:target_data_STM}
\end{equation}
\begin{equation}
  v_{k,D}^{\rm PC}
  =
  \sum_{j=0}^{D}
  {\rm bi}_{k-D}\ \ ({\rm mod}\ 2),
  \label{eq:target_data_PC}
\end{equation}
where ${\rm bi}_{k}$ is the $k$th random binary input data. 

\begin{figure*}
\centerline{\includegraphics[width=2.0\columnwidth]{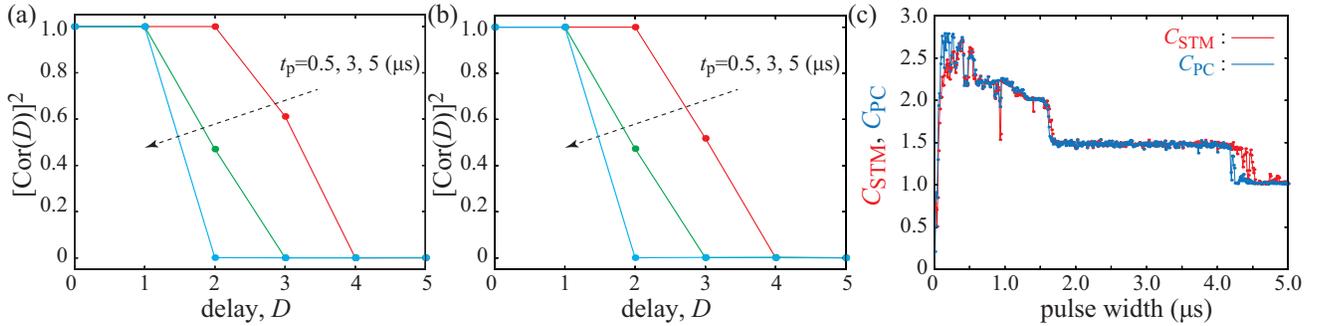}}
\caption{
         Squares of the correlation coefficients, for (a) the STM and (b) PC tasks at pulse widths of 0.5 (red), 3.0 (green) and 5.0 (blue) $\mu$s. 
         (c) Dependences of the STM (red) and PC (blue) capacities on pulse width, where $\nu=0.2$. 
         \vspace{-3ex}}
\label{fig:fig3}
\end{figure*}

After determining the weight, we apply different $N^{\prime}=1000$ random binary pulses as a test set 
and evaluate the reproducibility of the target output $v_{n,D}^{\prime}$ ($n=1,2,\cdots,N^{\prime}$) from the system output defined as 
\begin{equation}
  v_{{\rm R},n,D}^{\prime}
  =
  \sum_{i=1}^{N_{\rm node}+1}
  w_{D,i}
  s_{n,i}^{\prime}, 
\end{equation}
where $s_{n,i}^{\prime}$ is the vortex-core position in the presence of $N^{\prime}$ random pulses. 
Figure \ref{fig:fig2}(c) shows the target output (red) and the system output (blue) for the STM task with a pulse width of $3.0$ $\mu$s, 
where the delay varies as $D=1$, $2$ and $3$ from left to right. 
The system output well reproduces the target output when the delay is small, 
whereas the reproducibility becomes low as the delay increases. 
The reproducibility is quantitatively characterised by the correlation coefficient ${\rm Cor}(D)$ between the target output and system output 
defined as 
\begin{equation}
\begin{split}
  &
  {\rm Cor}(D)
  \equiv
  \frac{\sum_{n=1}^{N^{\prime}} \left(v_{n,D}^{\prime} - \langle v_{n,D}^{\prime} \rangle \right) \left( v_{R,n,D}^{\prime} - \langle v_{{\rm R},n,D}^{\prime} \rangle \right)}
    {\sqrt{\sum_{n=1}^{N^{\prime}}\left(v_{n,D}^{\prime} - \langle v_{n,D}^{\prime} \rangle \right)^{2}  \sum_{n=1}^{N^{\prime}}\left( v_{{\rm R},n,D}^{\prime} - \langle v_{{\rm R},n,D}^{\prime} \rangle \right)^{2}}},
  \label{eq:correlation}
\end{split}
\end{equation}
where $\langle \cdots \rangle$ is the averaged value. 
Figures \ref{fig:fig3}(a) and \ref{fig:fig3}(b) show the square of the correlation coefficients for the STM and PC tasks, respectively,
at pulse widths of $t_{\rm p}=0.5$, $3.0$ and $5.0$ $\mu$s. 
The correlation coefficient becomes small as the delay $D$ increases and becomes nearly zero as shown in Fig. \ref{fig:fig3}. 
This is also indicated in Fig. \ref{fig:fig2}(c) where the reproducibility of the input data decreases as the delay increases. 



The STM and PC capacities, denoted as $C_{\rm STM}$ and $C_{\rm PC}$, respectively, are defined as 
\begin{equation}
  C
  =
  \sum_{D=1}^{D_{\rm max}}
  \left[
    {\rm Cor}(D)
  \right]^{2}. 
\end{equation}
The maximum delay $D_{\rm max}$ in this work is set to $20$, which is sufficient to evaluate the saturated values of the capacities. 
We also note that the number of the training data, $N=1000$, is also sufficient to saturate the values of the capacities; see Supplemental Information. 
Figure \ref{fig:fig3}(c) shows the dependences of $C_{\rm STM}$ (red) and $C_{\rm PC}$ (blue) on the pulse width. 
It should be emphasised that both $C_{\rm STM}$ and $C_{\rm PC}$ show step-like behaviour with respect to the pulse width. 
When the pulse width is relatively short ($\lesssim 2.0$ $\mu$s), the capacities remain at large values over $2.0$. 
In a middle range of the pulse width ($2.0 \lesssim t_{\rm p} \lesssim 4.5$ $\mu$s), on the other hand, the capacities suddenly drop to $1.5$. 
The capacities again drop to $1.0$ when the pulse width becomes longer ($t_{\rm p} \gtrsim 4.5$ $\mu$s). 


\subsection*{Role of current-dependent relaxation on memory function}
\label{sec:Role of current-dependent relaxation on memory function}

The step-like behavior of the capacities in Fig. \ref{fig:fig3}(c) can be attributed to the current-dependent relaxation phenomenon of the vortex core. 
To explain this point, we first focus on an analytical solution of the Thiele equation. 
Neglecting the small spin-transfer torque due to the in-plane component of the reference layer's magnetisation 
and higher order terms of $s$ ($|s| \le 1$) and $\alpha$ ($\alpha \ll 1$), 
the Thiele equation for the vortex centre position $s$ becomes 
\begin{equation}
  \dot{s}
  =
  a s
  -
  b s^{3},
  \label{eq:Thiele_s}
\end{equation}
where $a$ and $b$ are defined as 
\begin{align}
  a
  =
  \frac{|\mathscr{D}| \kappa}{G^{2}}
  \left(
    \frac{Ga_{J}Jp_{z}}{|\mathscr{D}|\kappa}
    -
    1
  \right), 
&&
  b
  =
  \frac{|\mathscr{D}| \kappa}{G^{2}}
  \left(
    \xi
    +
    \zeta
  \right). 
\end{align}
Equation (\ref{eq:Thiele_s}) is the Stuart-Landau equation \cite{strogatz01,pikovsky03} for the real variable $s$. 
The vortex core is stabilised at the disc centre when $a<0$, 
whereas a limit-cycle oscillation appears when $a>0$. 
The solution of Eq. (\ref{eq:Thiele_s}) for $a>0$ with the initial condition $s(t=0)=s(0)$ is given by 
\begin{equation}
  s(t)
  =
  \frac{s(0) e^{at}}{\sqrt{1 + s(0)^{2} (b/a) (e^{2at}-1)}}.
  \label{eq:sol_s}
\end{equation}
The solution saturates to $\lim_{t \to \infty}s(t)=\sqrt{a/b}$. 
The normalised core position should satisfy $0 \le s \le 1$. 
Therefore, the condition to excite the limit cycle oscillation is $J_{\rm c1} < J < J_{\rm c2}$, 
where $J_{\rm c1}=(|\mathscr{D}|\kappa)/(Ga_{J}p_{z})$ and $J_{\rm c2}=J_{\rm c1}(1+\xi+\zeta)$ are determined by the conditions of $a>0$ and $a/b<1$, respectively. 
We note that the saturated core position depends on the disc radius $R$ through the parameter $\kappa \propto 1/R$. 
The parameter $\kappa$ determines the magnetic potential energy of the vortex core, as shown in Eq. (\ref{eq:magnetic_energy}), 
and becomes small as the disc radius $R$ increases because, in a large disc, a tiny displacement of the vortex core does not change the magnetic energy significantly. 
Since the spin-transfer torque should overcome the damping torque, which helps keep the vortex core close to the disc centre to minimise the magnetic energy, in driving the vortex-core dynamics, 
the critical current density $J_{\rm c1}$ is proportional to the parameter $\kappa$. 
Accordingly, the saturated core position $\sqrt{a/b}$ at a given current density $J$ depends on the disc radius $R$ through the parameter $\kappa$ in the critical current density. 
Roughly speaking, the saturated core position becomes large as the disc radius $R$ increases because the critical current density becomes small for a large $R$. 


Equation (\ref{eq:Thiele_s}) indicates that the time scale for the exponential evolution of the vortex core, given by $1/a$, depends on the input data through the current density; 
a large current density results in a fast relaxation. 
For example, $1/a$ for the current density $J$ with ${\rm bi}=0$ is $0.36$ $\mu$s for the present parameters, whereas it becomes $0.14$ $\mu$s when ${\rm bi}=1$.


Before proceeding to the discussion, we give a brief comment on the relation between the parameter $1/a$ and the time scale for reservoir computing. 
Reservoir computing estimates the information of the past input from the temporal response of the reservoir. 
Therefore, the relaxation phenomenon is key to performing physical reservoir computing. 
The parameter $1/a(\simeq 0.36\ {\rm or}\ 0.14\ \mu{\rm s})$ characterises the exponential relaxation of the vortex core. 
We, however, note that nearly $e^{-1}\simeq0.37$, i.e., $37\%$ of the relaxation process to the saturated value $\sqrt{a/b}$ is not completed even after a time scale of $1/a$ has passed. 
The vortex dynamics, such as those at the last part of the relaxation process, can also be used in reservoir computing. 
As a result, the time scale of the relaxation dynamics of the vortex core applicable to reservoir computing is longer than $1/a$. 
For example, the pulse width ($\sim 4.5$ $\mu$s) at which the capacities drop from $1.5$ to $1.0$ shown in Fig. \ref{fig:fig3}(c) is not exactly the same with the parameter $1/a$, 
although such a drop of the capacities is related to the relaxation phenomenon of the vortex core, as explained below. 
Therefore, we use the word "relaxation time" in the following as a time scale to determine the performance of reservoir computing. 
However, defining the relaxation time quantitatively is difficult because it depends on the number of significant figures in the calculations. 
We also note that the relaxation time depends on the current magnitude, as in the case of the parameter $1/a$, because of the current-dependent relaxation dynamics. 
If the relaxation time is faster than the pulse width, the system rapidly relaxes to a steady state and loses the history of the time-series data. 
In other words, the pulse width should be shorter than the relaxation time of the reservoir. 
An excessively short pulse width is, however, also not preferable to reservoir computing 
because the change in dynamical response with respect to a pulse input becomes small for a short pulse width. 




\begin{figure*}
\centerline{\includegraphics[width=2.0\columnwidth]{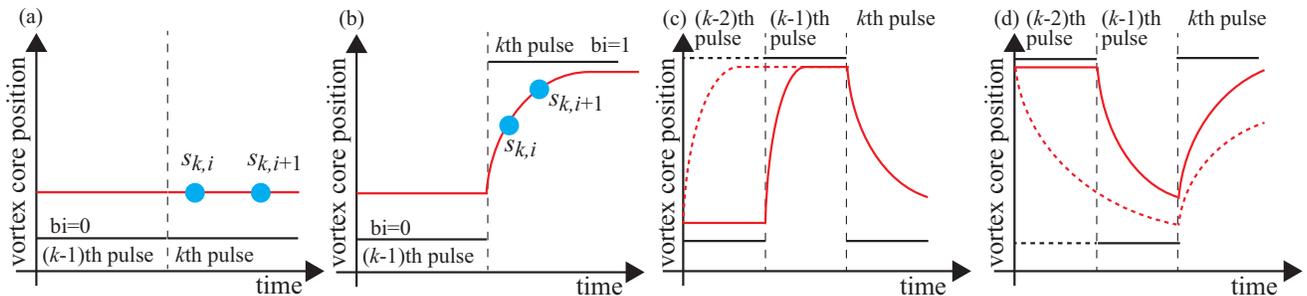}}
\caption{
        (a), (b)Schematic pictures of the vortex dynamics (red) in the presence of 
        two ($(k-1)$th and $k$th) pulses (black), where the $k$th and $(k-1)$th pulses are the same in (a) and different in (b). 
        The blue circles represent the nodes. 
        (c), (d), Schematic views of similar dynamics with three pulses, 
        where the solid and dotted lines show the cases when the values of the $(k-2)$th and $(k-1)$th binary pulses are different and the same. 
         \vspace{-3ex}}
\label{fig:fig4}
\end{figure*}




Regarding these points, the identification of the data inputted one sequence before the present pulse can be explained as follows. 
As an example, let us assume that the vortex core remains in a steady state under a binary pulse of ${\rm bi}=0$. 
For convenience, we name this pulse as the $(k-1)$th pulse. 
If the next ($k$th) pulse is ${\rm bi}=0$, the vortex state is unchanged, as shown in Fig. \ref{fig:fig4}(a). 
On the other hand, the vortex core changes the position if the next pulse is ${\rm bi}=1$, as shown in Fig. \ref{fig:fig4}(b). 
In other words, the response with respect to the $k$th pulse is affected by the $(k-1)$th pulse. 
Therefore, the value of the $(k-1)$th pulse can be estimated from the response under the $k$th pulse, corresponding to $[{\rm Cor}(D=1)]^{2}= 1$. 
This fact results in capacities of $1.0$ in a long pulse-width limit shown in Fig. \ref{fig:fig3}(c). 


Another saturated value, $1.5$, of the capacities in the middle pulse-width range is explained in a similar way, 
by taking into account the current-dependent relaxation time. 
Let us consider the injection of three pulses, named as the $(k-2)$th, $(k-1)$th and $k$th pulses, as shown in Figs. \ref{fig:fig4}(c) and \ref{fig:fig4}(d). 
The solid and dotted lines in the figures show the vortex dynamics when the $(k-2)$th and $(k-1)$th binary pulses are different and the same, respectively. 
In Fig. \ref{fig:fig4}(c), the $(k-1)$th pulse is ${\rm bi}=1$. 
In this case, the vortex core saturates rapidly, and thus, the vortex core is in a steady state at the end of the $(k-1)$th pulse, independent of the $(k-2)$th pulse. 
This fact indicates that the $(k-2)$th input data cannot be identified from the vortex dynamics under the presence of the $k$th input. 
On the other hand, when the $(k-1)$ th pulse is ${\rm bi}=0$, the vortex dynamics under the presence of the $k$th pulse depends on the $(k-2)$th pulse. 
This is because the relaxation time is slow, and thus, the vortex state at the end of the $(k-1)$th pulse reflects the state under the $(k-2)$th pulse, as shown in Fig. \ref{fig:fig4}(d). 
As a result, the $(k-2)$th input can be identified when the $(k-1)$th pulse is ${\rm bi}=0$. 
Therefore, the reproducibility of the $(k-2)$th data is $50\%$, resulting in $[{\rm Cor}(D=2)]^{2}= 0.5$ and the capacity value of $1.5$. 
As can be seen in this explanation, the step-like behaviour of the memory function in Fig. \ref{fig:fig3}(c) can be attributed to the current-dependent relaxation mechanism of the vortex dynamics. 


\subsection*{Memory capacity for different input strength}

\begin{figure*}
\centerline{\includegraphics[width=2.0\columnwidth]{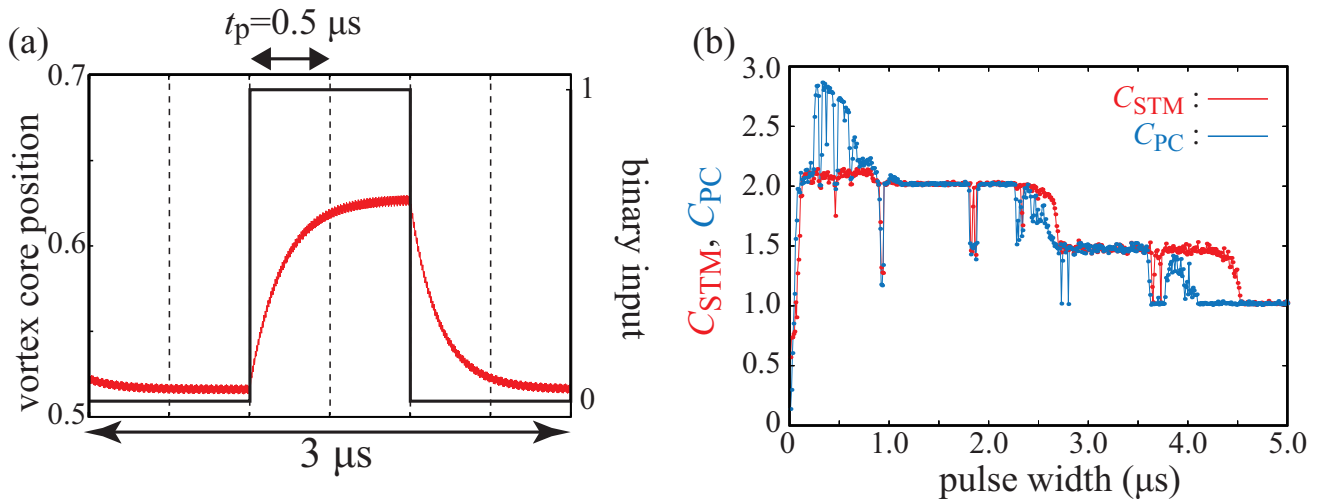}}
\caption{
         (a) Time evolutions of the vortex-core position (red) and the binary pulses (black) with widths of $0.5$ $\mu$s and 
         (b) dependences of the STM (red) and PC (blue) capacities on pulse width, 
         where the dimensionless parameter $\nu$ that determinines the difference in magnitudes of the binary inputs is $0.05$. 
         \vspace{-3ex}}
\label{fig:fig5}
\end{figure*}

The role of the relaxation time of the vortex core on the memory function for reservoir computing can be discussed from a different viewpoint. 
Figure \ref{fig:fig5}(a) shows the dynamics of the centre position of the vortex core in the presence of six pulses with widths of $0.5$ $\mu$s. 
The situation is the same with that shown in Fig. \ref{fig:fig3}(a), but here, we use a relatively weak input pulse characterised by $\nu=0.05$. 
Because of the small input-pulse strength $\nu$, the change of $s(t)$ with respect to the input current is small compared with that shown in Fig. \ref{fig:fig2}(a). 



Figure \ref{fig:fig5}(b) shows the dependences of the STM and PC capacities on the pulse width for $\nu=0.05$. 
Note that we argue above that the value of $1.5$ in the memory capacity reflects the difference in the relaxation times between two states. 
Due to the small parameter $\nu$, the relaxation time from the ${\rm bi}=0$ to ${\rm bi}=1$ state becomes close to that of the opposite case, 
compared with the system studied in the previous sections. 
As a result, the reservoir can distinguish both signals in a relatively large pulse-width range. 
Therefore, the range corresponding to the value of $C=2$ for the memory capacities is enlarged, compared with that found in Fig. \ref{fig:fig3}(c) in the main text, 
and the step-like behavior is observed even in a short pulse-width range. 
In addition, the pulse width at which the jump from $C=2$ to $C=1.5$ occurs is shifted to a long pulse-width region. 
The results also indicate that the step-like behaviour originates from the current-dependent vortex relation, which is controlled by the parameter $\nu$. 
In Supplemental Information, we also provide the data showing the STM and PC capacities for various values of $\nu$, 
which also support the argument here. 


\section*{Discussion}

In summary, physical reservoir computing using magnetic vortex-core dynamics in a fine-structured ferromagnet was performed by solving the Thiele equation numerically. 
The step-like dependence of the STM and PC capacities on the pulse width was found, 
where the capacities remain at a value of $1.5$ for a certain range of the pulse width, and drop to $1.0$ for a long pulse-width limit. 
Such a half-integer memory capacity originated from the current-dependent relaxation mechanism of the vortex core, 
where a fast relaxation caused by a large input led to a fast fading of the input memory, 
whereas a slow relaxation by a small input enabled the reservoir to keep the input memory for a relatively long time.
Using a small input-pulse made the difference of the relaxation times between the two states small and suppressed such a fast fading memory. 
Accordingly, the input-pulse range corresponding to a relatively large capacity, $2.0$, was enlarged. 

In practical applications, the input data for reservoir computing is converted to electrical inputs by the data conversion, and a preprocessing method is applied to the data \cite{tsunegi18}. 
The present work indicates that an appropriate choice of pulse width is necessary in the data conversion to achieve high performance of physical reservoir computing. 
The value of the appropriate pulse width relates to the relaxation time of the reservoir. 
Spintronics technology provides various kinds of structures, such as macrospin, domain wall, spin-wave and/or skyrmions, by arranging the materials, designs and topologies. 
The process of the data conversion will depend on the devices because the relaxation time depends on the structures. 
This study therefore provides a crucial guideline for such device design. 


\section*{Methods}

The Thiele equation is solved by the fourth-order Runge-Kutta method. 
The minimisation of the error between the system output and the target data is performed using the Moore-Penrose pseudo inverse matrix determined by the singular value decomposition. 



\section*{Acknowledgements}

The authors acknowledge Shinji Miwa and Takehiko Yorozu for their valuable discussions. 
This paper was based on the results obtained from a project (Innovative AI Chips and Next-Generation Computing Technology Development/(2) 
Development of next-generation computing technologies/Exploration of Neuromorphic Dynamics towards Future Symbiotic Society) commissioned by NEDO. 


\section*{Author contributions statement}

K.N., S.T., T.T., and H.K. designed the project. 
T.Y., N.A., K.N., and T.T. directed the investigation. 
T.Y. and T.T. developed the program codes, performed the simulations, prepared the figures, and wrote the manuscript. 
All authors contributed to discussing the results. 


\section*{Competing interests}

The authors declare no competing interests. 


\section*{Additional information}

\textbf{Correspondence} and requests for materials should be addressed to T.T.
\\
\textbf{Supplementary information} accompanies this paper. 




%


\begin{thebibliography}{41}%
\makeatletter
\providecommand \@ifxundefined [1]{%
 \@ifx{#1\undefined}
}%
\providecommand \@ifnum [1]{%
 \ifnum #1\expandafter \@firstoftwo
 \else \expandafter \@secondoftwo
 \fi
}%
\providecommand \@ifx [1]{%
 \ifx #1\expandafter \@firstoftwo
 \else \expandafter \@secondoftwo
 \fi
}%
\providecommand \natexlab [1]{#1}%
\providecommand \enquote  [1]{``#1''}%
\providecommand \bibnamefont  [1]{#1}%
\providecommand \bibfnamefont [1]{#1}%
\providecommand \citenamefont [1]{#1}%
\providecommand \href@noop [0]{\@secondoftwo}%
\providecommand \href [0]{\begingroup \@sanitize@url \@href}%
\providecommand \@href[1]{\@@startlink{#1}\@@href}%
\providecommand \@@href[1]{\endgroup#1\@@endlink}%
\providecommand \@sanitize@url [0]{\catcode `\\12\catcode `\$12\catcode
  `\&12\catcode `\#12\catcode `\^12\catcode `\_12\catcode `\%12\relax}%
\providecommand \@@startlink[1]{}%
\providecommand \@@endlink[0]{}%
\providecommand \url  [0]{\begingroup\@sanitize@url \@url }%
\providecommand \@url [1]{\endgroup\@href {#1}{\urlprefix }}%
\providecommand \urlprefix  [0]{URL }%
\providecommand \Eprint [0]{\href }%
\providecommand \doibase [0]{https://doi.org/}%
\providecommand \selectlanguage [0]{\@gobble}%
\providecommand \bibinfo  [0]{\@secondoftwo}%
\providecommand \bibfield  [0]{\@secondoftwo}%
\providecommand \translation [1]{[#1]}%
\providecommand \BibitemOpen [0]{}%
\providecommand \bibitemStop [0]{}%
\providecommand \bibitemNoStop [0]{.\EOS\space}%
\providecommand \EOS [0]{\spacefactor3000\relax}%
\providecommand \BibitemShut  [1]{\csname bibitem#1\endcsname}%
\let\auto@bib@innerbib\@empty
\bibitem [{\citenamefont {Mandic}\ and\ \citenamefont
  {Chambers}(2001)}]{mandic01}%
  \BibitemOpen
  \bibfield  {author} {\bibinfo {author} {\bibfnamefont {D.~P.}\ \bibnamefont
  {Mandic}}\ and\ \bibinfo {author} {\bibfnamefont {J.~A.}\ \bibnamefont
  {Chambers}},\ }\href@noop {} {\emph {\bibinfo {title} {Recurrent neural
  networks for prediction: learning algorithms, architectures and stability}}}\
  (\bibinfo  {publisher} {Wiley},\ \bibinfo {year} {2001})\BibitemShut
  {NoStop}%
\bibitem [{\citenamefont {Maas}\ \emph {et~al.}(2002)\citenamefont {Maas},
  \citenamefont {Natschl\"ager},\ and\ \citenamefont {Markram}}]{maas02}%
  \BibitemOpen
  \bibfield  {author} {\bibinfo {author} {\bibfnamefont {W.}~\bibnamefont
  {Maas}}, \bibinfo {author} {\bibfnamefont {T.}~\bibnamefont
  {Natschl\"ager}},\ and\ \bibinfo {author} {\bibfnamefont {H.}~\bibnamefont
  {Markram}},\ }\bibfield  {title} {\bibinfo {title} {Real-{Time} {Computing}
  {Without} {Stable} {States}: {A} {New} {Framework} for {Neural} {Computation}
  {Based} on {Perturbations}},\ }\href@noop {} {\bibfield  {journal} {\bibinfo
  {journal} {Neural Comput.}\ }\textbf {\bibinfo {volume} {14}},\ \bibinfo
  {pages} {2531} (\bibinfo {year} {2002})}\BibitemShut {NoStop}%
\bibitem [{\citenamefont {Jaeger}\ and\ \citenamefont {Haas}(2004)}]{jaeger04}%
  \BibitemOpen
  \bibfield  {author} {\bibinfo {author} {\bibfnamefont {H.}~\bibnamefont
  {Jaeger}}\ and\ \bibinfo {author} {\bibfnamefont {H.}~\bibnamefont {Haas}},\
  }\bibfield  {title} {\bibinfo {title} {Harnessing {Nonlinearity}:
  {Predicting} {Chaotic} {Systems} and {Saving} {Energy} in {Wireless}
  {Communication}},\ }\href@noop {} {\bibfield  {journal} {\bibinfo  {journal}
  {Science}\ }\textbf {\bibinfo {volume} {304}},\ \bibinfo {pages} {78}
  (\bibinfo {year} {2004})}\BibitemShut {NoStop}%
\bibitem [{\citenamefont {Verstraeten}\ \emph {et~al.}(2007)\citenamefont
  {Verstraeten}, \citenamefont {Schrauwen}, \citenamefont {D'Haene},\ and\
  \citenamefont {Stroobandt}}]{verstraeten07}%
  \BibitemOpen
  \bibfield  {author} {\bibinfo {author} {\bibfnamefont {D.}~\bibnamefont
  {Verstraeten}}, \bibinfo {author} {\bibfnamefont {B.}~\bibnamefont
  {Schrauwen}}, \bibinfo {author} {\bibfnamefont {M.}~\bibnamefont {D'Haene}},\
  and\ \bibinfo {author} {\bibfnamefont {D.}~\bibnamefont {Stroobandt}},\
  }\bibfield  {title} {\bibinfo {title} {An experimental unification of
  reservoir computing methods},\ }\href@noop {} {\bibfield  {journal} {\bibinfo
   {journal} {Neural Netw.}\ }\textbf {\bibinfo {volume} {20}},\ \bibinfo
  {pages} {391} (\bibinfo {year} {2007})}\BibitemShut {NoStop}%
\bibitem [{\citenamefont {Hermans}\ and\ \citenamefont
  {Schrauwen}(2010)}]{hermans10}%
  \BibitemOpen
  \bibfield  {author} {\bibinfo {author} {\bibfnamefont {M.}~\bibnamefont
  {Hermans}}\ and\ \bibinfo {author} {\bibfnamefont {B.}~\bibnamefont
  {Schrauwen}},\ }\bibfield  {title} {\bibinfo {title} {Memory in linear
  recurrent neural networks in continuous time},\ }\href@noop {} {\bibfield
  {journal} {\bibinfo  {journal} {Neural Netw.}\ }\textbf {\bibinfo {volume}
  {23}},\ \bibinfo {pages} {341} (\bibinfo {year} {2010})}\BibitemShut
  {NoStop}%
\bibitem [{\citenamefont {Appeltant}\ \emph {et~al.}(2011)\citenamefont
  {Appeltant}, \citenamefont {Soriano}, \citenamefont {der Sande},
  \citenamefont {Danckaert}, \citenamefont {Massar}, \citenamefont {Dambre},
  \citenamefont {Schrauwen}, \citenamefont {Mirasso},\ and\ \citenamefont
  {Fischer}}]{appeltant11}%
  \BibitemOpen
  \bibfield  {author} {\bibinfo {author} {\bibfnamefont {L.}~\bibnamefont
  {Appeltant}}, \bibinfo {author} {\bibfnamefont {M.~C.}\ \bibnamefont
  {Soriano}}, \bibinfo {author} {\bibfnamefont {G.~V.}\ \bibnamefont {der
  Sande}}, \bibinfo {author} {\bibfnamefont {J.}~\bibnamefont {Danckaert}},
  \bibinfo {author} {\bibfnamefont {S.}~\bibnamefont {Massar}}, \bibinfo
  {author} {\bibfnamefont {J.}~\bibnamefont {Dambre}}, \bibinfo {author}
  {\bibfnamefont {B.}~\bibnamefont {Schrauwen}}, \bibinfo {author}
  {\bibfnamefont {C.~R.}\ \bibnamefont {Mirasso}},\ and\ \bibinfo {author}
  {\bibfnamefont {I.}~\bibnamefont {Fischer}},\ }\bibfield  {title} {\bibinfo
  {title} {Information processing using a single dynamical node as complex
  system},\ }\href@noop {} {\bibfield  {journal} {\bibinfo  {journal} {Nat.
  Commun.}\ }\textbf {\bibinfo {volume} {2}},\ \bibinfo {pages} {468} (\bibinfo
  {year} {2011})}\BibitemShut {NoStop}%
\bibitem [{\citenamefont {Grigoryeva}\ and\ \citenamefont
  {Ortega}(2018)}]{grigoryeva18}%
  \BibitemOpen
  \bibfield  {author} {\bibinfo {author} {\bibfnamefont {L.}~\bibnamefont
  {Grigoryeva}}\ and\ \bibinfo {author} {\bibfnamefont {J.-P.}\ \bibnamefont
  {Ortega}},\ }\bibfield  {title} {\bibinfo {title} {Echo state networks are
  universal},\ }\href@noop {} {\bibfield  {journal} {\bibinfo  {journal}
  {Neural Netw.}\ }\textbf {\bibinfo {volume} {108}},\ \bibinfo {pages} {495}
  (\bibinfo {year} {2018})}\BibitemShut {NoStop}%
\bibitem [{\citenamefont {R\"ohm}\ and\ \citenamefont
  {L\"udge}(2018)}]{rohm18}%
  \BibitemOpen
  \bibfield  {author} {\bibinfo {author} {\bibfnamefont {A.}~\bibnamefont
  {R\"ohm}}\ and\ \bibinfo {author} {\bibfnamefont {K.}~\bibnamefont
  {L\"udge}},\ }\bibfield  {title} {\bibinfo {title} {Multiplexed networks:
  reservoir computing with virtual and real nodes},\ }\href@noop {} {\bibfield
  {journal} {\bibinfo  {journal} {J. Phys. Commun.}\ }\textbf {\bibinfo
  {volume} {2}},\ \bibinfo {pages} {085007} (\bibinfo {year}
  {2018})}\BibitemShut {NoStop}%
\bibitem [{\citenamefont {Nakajima}(2020)}]{nakajima20}%
  \BibitemOpen
  \bibfield  {author} {\bibinfo {author} {\bibfnamefont {K.}~\bibnamefont
  {Nakajima}},\ }\bibfield  {title} {\bibinfo {title} {Physical reservoir
  computing - an introductory perspective},\ }\href@noop {} {\bibfield
  {journal} {\bibinfo  {journal} {Jpn. J. Appl. Phys.}\ }\textbf {\bibinfo
  {volume} {59}},\ \bibinfo {pages} {060501} (\bibinfo {year}
  {2020})}\BibitemShut {NoStop}%
\bibitem [{\citenamefont {Brunner}\ \emph {et~al.}(2013)\citenamefont
  {Brunner}, \citenamefont {Soriano}, \citenamefont {Mirasso},\ and\
  \citenamefont {Fischer}}]{brunner13}%
  \BibitemOpen
  \bibfield  {author} {\bibinfo {author} {\bibfnamefont {D.}~\bibnamefont
  {Brunner}}, \bibinfo {author} {\bibfnamefont {M.~C.}\ \bibnamefont
  {Soriano}}, \bibinfo {author} {\bibfnamefont {C.~R.}\ \bibnamefont
  {Mirasso}},\ and\ \bibinfo {author} {\bibfnamefont {I.}~\bibnamefont
  {Fischer}},\ }\bibfield  {title} {\bibinfo {title} {Parallel photonic
  information processing at gigabyte per second data rates using trasient
  states},\ }\href@noop {} {\bibfield  {journal} {\bibinfo  {journal} {Nat.
  Commun.}\ }\textbf {\bibinfo {volume} {4}},\ \bibinfo {pages} {1364}
  (\bibinfo {year} {2013})}\BibitemShut {NoStop}%
\bibitem [{\citenamefont {Nakajima}\ \emph {et~al.}(2015)\citenamefont
  {Nakajima}, \citenamefont {Hauser}, \citenamefont {Li},\ and\ \citenamefont
  {Pfeifer}}]{nakajima15}%
  \BibitemOpen
  \bibfield  {author} {\bibinfo {author} {\bibfnamefont {K.}~\bibnamefont
  {Nakajima}}, \bibinfo {author} {\bibfnamefont {H.}~\bibnamefont {Hauser}},
  \bibinfo {author} {\bibfnamefont {T.}~\bibnamefont {Li}},\ and\ \bibinfo
  {author} {\bibfnamefont {R.}~\bibnamefont {Pfeifer}},\ }\bibfield  {title}
  {\bibinfo {title} {Information processing via physical soft body},\
  }\href@noop {} {\bibfield  {journal} {\bibinfo  {journal} {Sci. Rep.}\
  }\textbf {\bibinfo {volume} {5}},\ \bibinfo {pages} {10487} (\bibinfo {year}
  {2015})}\BibitemShut {NoStop}%
\bibitem [{\citenamefont {Fujii}\ and\ \citenamefont
  {Nakajima}(2017)}]{fujii17}%
  \BibitemOpen
  \bibfield  {author} {\bibinfo {author} {\bibfnamefont {K.}~\bibnamefont
  {Fujii}}\ and\ \bibinfo {author} {\bibfnamefont {K.}~\bibnamefont
  {Nakajima}},\ }\bibfield  {title} {\bibinfo {title} {Harnessing
  {Disordered}-{Ensemble} {Quantum} {Dynamics} for {Machine} {Learning}},\
  }\href@noop {} {\bibfield  {journal} {\bibinfo  {journal} {Phys. Rev.
  Applied}\ }\textbf {\bibinfo {volume} {8}},\ \bibinfo {pages} {024030}
  (\bibinfo {year} {2017})}\BibitemShut {NoStop}%
\bibitem [{\citenamefont {Dion}\ \emph {et~al.}(2018)\citenamefont {Dion},
  \citenamefont {Mejaouri},\ and\ \citenamefont {Sylvestre}}]{dion18}%
  \BibitemOpen
  \bibfield  {author} {\bibinfo {author} {\bibfnamefont {G.}~\bibnamefont
  {Dion}}, \bibinfo {author} {\bibfnamefont {S.}~\bibnamefont {Mejaouri}},\
  and\ \bibinfo {author} {\bibfnamefont {J.}~\bibnamefont {Sylvestre}},\
  }\bibfield  {title} {\bibinfo {title} {Reservoir computing with a single
  delay-coupled non-linear mechanical oscillator},\ }\href@noop {} {\bibfield
  {journal} {\bibinfo  {journal} {J. Appl. Phys.}\ }\textbf {\bibinfo {volume}
  {124}},\ \bibinfo {pages} {152132} (\bibinfo {year} {2018})}\BibitemShut
  {NoStop}%
\bibitem [{\citenamefont {Nakajima}\ \emph {et~al.}(2019)\citenamefont
  {Nakajima}, \citenamefont {Fujii}, \citenamefont {Negoro}, \citenamefont
  {Mitarai},\ and\ \citenamefont {Kitagawa}}]{nakajima19}%
  \BibitemOpen
  \bibfield  {author} {\bibinfo {author} {\bibfnamefont {K.}~\bibnamefont
  {Nakajima}}, \bibinfo {author} {\bibfnamefont {K.}~\bibnamefont {Fujii}},
  \bibinfo {author} {\bibfnamefont {M.}~\bibnamefont {Negoro}}, \bibinfo
  {author} {\bibfnamefont {K.}~\bibnamefont {Mitarai}},\ and\ \bibinfo {author}
  {\bibfnamefont {M.}~\bibnamefont {Kitagawa}},\ }\bibfield  {title} {\bibinfo
  {title} {Boosting {Computational} {Power} through {Spatial} {Multiplexing} in
  {Quantum} {Reservoir} {Computing}},\ }\href@noop {} {\bibfield  {journal}
  {\bibinfo  {journal} {Phys. Rev. Applied}\ }\textbf {\bibinfo {volume}
  {11}},\ \bibinfo {pages} {034021} (\bibinfo {year} {2019})}\BibitemShut
  {NoStop}%
\bibitem [{\citenamefont {Torrejon}\ \emph {et~al.}(2017)\citenamefont
  {Torrejon}, \citenamefont {Riou}, \citenamefont {Araujo}, \citenamefont
  {Tsunegi}, \citenamefont {Khalsa}, \citenamefont {Querlioz}, \citenamefont
  {Bortolotti}, \citenamefont {Cros}, \citenamefont {Yakushiji}, \citenamefont
  {Fukushima}, \citenamefont {Kubota}, \citenamefont {Yuasa}, \citenamefont
  {Stiles},\ and\ \citenamefont {Grollier}}]{torrejon17}%
  \BibitemOpen
  \bibfield  {author} {\bibinfo {author} {\bibfnamefont {J.}~\bibnamefont
  {Torrejon}}, \bibinfo {author} {\bibfnamefont {M.}~\bibnamefont {Riou}},
  \bibinfo {author} {\bibfnamefont {F.~A.}\ \bibnamefont {Araujo}}, \bibinfo
  {author} {\bibfnamefont {S.}~\bibnamefont {Tsunegi}}, \bibinfo {author}
  {\bibfnamefont {G.}~\bibnamefont {Khalsa}}, \bibinfo {author} {\bibfnamefont
  {D.}~\bibnamefont {Querlioz}}, \bibinfo {author} {\bibfnamefont
  {P.}~\bibnamefont {Bortolotti}}, \bibinfo {author} {\bibfnamefont
  {V.}~\bibnamefont {Cros}}, \bibinfo {author} {\bibfnamefont {K.}~\bibnamefont
  {Yakushiji}}, \bibinfo {author} {\bibfnamefont {A.}~\bibnamefont
  {Fukushima}}, \bibinfo {author} {\bibfnamefont {H.}~\bibnamefont {Kubota}},
  \bibinfo {author} {\bibfnamefont {S.}~\bibnamefont {Yuasa}}, \bibinfo
  {author} {\bibfnamefont {M.~D.}\ \bibnamefont {Stiles}},\ and\ \bibinfo
  {author} {\bibfnamefont {J.}~\bibnamefont {Grollier}},\ }\bibfield  {title}
  {\bibinfo {title} {Neuromorphic computing with nanoscale spintronic
  oscillators},\ }\href@noop {} {\bibfield  {journal} {\bibinfo  {journal}
  {Nature}\ }\textbf {\bibinfo {volume} {547}},\ \bibinfo {pages} {428}
  (\bibinfo {year} {2017})}\BibitemShut {NoStop}%
\bibitem [{\citenamefont {Furuta}\ \emph {et~al.}(2018)\citenamefont {Furuta},
  \citenamefont {Fujii}, \citenamefont {Nakajima}, \citenamefont {Tsunegi},
  \citenamefont {Kubota}, \citenamefont {Suzuki},\ and\ \citenamefont
  {Miwa}}]{furuta18}%
  \BibitemOpen
  \bibfield  {author} {\bibinfo {author} {\bibfnamefont {T.}~\bibnamefont
  {Furuta}}, \bibinfo {author} {\bibfnamefont {K.}~\bibnamefont {Fujii}},
  \bibinfo {author} {\bibfnamefont {K.}~\bibnamefont {Nakajima}}, \bibinfo
  {author} {\bibfnamefont {S.}~\bibnamefont {Tsunegi}}, \bibinfo {author}
  {\bibfnamefont {H.}~\bibnamefont {Kubota}}, \bibinfo {author} {\bibfnamefont
  {Y.}~\bibnamefont {Suzuki}},\ and\ \bibinfo {author} {\bibfnamefont
  {S.}~\bibnamefont {Miwa}},\ }\bibfield  {title} {\bibinfo {title}
  {Macromagnetic {Simulation} for {Reservoir} {Computing} {Utilizing} {Spin}
  {Dynamics} in {Magnetic} {Tunnel} {Junctions}},\ }\href@noop {} {\bibfield
  {journal} {\bibinfo  {journal} {Phys. Rev. Applied}\ }\textbf {\bibinfo
  {volume} {10}},\ \bibinfo {pages} {034063} (\bibinfo {year}
  {2018})}\BibitemShut {NoStop}%
\bibitem [{\citenamefont {Tsunegi}\ \emph {et~al.}(2018)\citenamefont
  {Tsunegi}, \citenamefont {Taniguchi}, \citenamefont {Miwa}, \citenamefont
  {Nakajima}, \citenamefont {Yakusjiji}, \citenamefont {Fukushima},
  \citenamefont {Yuasa},\ and\ \citenamefont {Kubota}}]{tsunegi18}%
  \BibitemOpen
  \bibfield  {author} {\bibinfo {author} {\bibfnamefont {S.}~\bibnamefont
  {Tsunegi}}, \bibinfo {author} {\bibfnamefont {T.}~\bibnamefont {Taniguchi}},
  \bibinfo {author} {\bibfnamefont {S.}~\bibnamefont {Miwa}}, \bibinfo {author}
  {\bibfnamefont {K.}~\bibnamefont {Nakajima}}, \bibinfo {author}
  {\bibfnamefont {K.}~\bibnamefont {Yakusjiji}}, \bibinfo {author}
  {\bibfnamefont {A.}~\bibnamefont {Fukushima}}, \bibinfo {author}
  {\bibfnamefont {S.}~\bibnamefont {Yuasa}},\ and\ \bibinfo {author}
  {\bibfnamefont {H.}~\bibnamefont {Kubota}},\ }\bibfield  {title} {\bibinfo
  {title} {Evaluation of memory capacity of spin torque oscillator for
  recurrent neural networks},\ }\href@noop {} {\bibfield  {journal} {\bibinfo
  {journal} {Jpn. J. Appl. Phys.}\ }\textbf {\bibinfo {volume} {57}},\ \bibinfo
  {pages} {120307} (\bibinfo {year} {2018})}\BibitemShut {NoStop}%
\bibitem [{\citenamefont {Bourianoff}\ \emph {et~al.}(2018)\citenamefont
  {Bourianoff}, \citenamefont {Pinna}, \citenamefont {Sitte},\ and\
  \citenamefont {Everschor-Sitte}}]{bourianoff18}%
  \BibitemOpen
  \bibfield  {author} {\bibinfo {author} {\bibfnamefont {G.}~\bibnamefont
  {Bourianoff}}, \bibinfo {author} {\bibfnamefont {D.}~\bibnamefont {Pinna}},
  \bibinfo {author} {\bibfnamefont {M.}~\bibnamefont {Sitte}},\ and\ \bibinfo
  {author} {\bibfnamefont {K.}~\bibnamefont {Everschor-Sitte}},\ }\bibfield
  {title} {\bibinfo {title} {Potential implementation of reservoir computing
  models based on magnetic skyrmions},\ }\href@noop {} {\bibfield  {journal}
  {\bibinfo  {journal} {AIP Adv.}\ }\textbf {\bibinfo {volume} {8}},\ \bibinfo
  {pages} {055602} (\bibinfo {year} {2018})}\BibitemShut {NoStop}%
\bibitem [{\citenamefont {Nakane}\ \emph {et~al.}(2018)\citenamefont {Nakane},
  \citenamefont {Tanaka},\ and\ \citenamefont {Hirose}}]{nakane18}%
  \BibitemOpen
  \bibfield  {author} {\bibinfo {author} {\bibfnamefont {R.}~\bibnamefont
  {Nakane}}, \bibinfo {author} {\bibfnamefont {G.}~\bibnamefont {Tanaka}},\
  and\ \bibinfo {author} {\bibfnamefont {A.}~\bibnamefont {Hirose}},\
  }\bibfield  {title} {\bibinfo {title} {Reservoir {Computing} {With} {Spin}
  {Waves} {Excited} in a {Garnet} {Film}},\ }\href@noop {} {\bibfield
  {journal} {\bibinfo  {journal} {IEEE Access}\ }\textbf {\bibinfo {volume}
  {6}},\ \bibinfo {pages} {4462} (\bibinfo {year} {2018})}\BibitemShut
  {NoStop}%
\bibitem [{\citenamefont {Nomura}\ \emph {et~al.}(2019)\citenamefont {Nomura},
  \citenamefont {Furuta}, \citenamefont {Tsujimoto}, \citenamefont
  {Kuwabiraki}, \citenamefont {Peper}, \citenamefont {Tamura}, \citenamefont
  {Miwa}, \citenamefont {Goto}, \citenamefont {Nakatani},\ and\ \citenamefont
  {Suzuki}}]{nomura19}%
  \BibitemOpen
  \bibfield  {author} {\bibinfo {author} {\bibfnamefont {H.}~\bibnamefont
  {Nomura}}, \bibinfo {author} {\bibfnamefont {T.}~\bibnamefont {Furuta}},
  \bibinfo {author} {\bibfnamefont {K.}~\bibnamefont {Tsujimoto}}, \bibinfo
  {author} {\bibfnamefont {Y.}~\bibnamefont {Kuwabiraki}}, \bibinfo {author}
  {\bibfnamefont {F.}~\bibnamefont {Peper}}, \bibinfo {author} {\bibfnamefont
  {E.}~\bibnamefont {Tamura}}, \bibinfo {author} {\bibfnamefont
  {S.}~\bibnamefont {Miwa}}, \bibinfo {author} {\bibfnamefont {M.}~\bibnamefont
  {Goto}}, \bibinfo {author} {\bibfnamefont {R.}~\bibnamefont {Nakatani}},\
  and\ \bibinfo {author} {\bibfnamefont {Y.}~\bibnamefont {Suzuki}},\
  }\bibfield  {title} {\bibinfo {title} {Reservoir computing with
  dipole-coupled nanomagnets},\ }\href@noop {} {\bibfield  {journal} {\bibinfo
  {journal} {Jpn. J. Appl. Phys.}\ }\textbf {\bibinfo {volume} {58}},\ \bibinfo
  {pages} {070901} (\bibinfo {year} {2019})}\BibitemShut {NoStop}%
\bibitem [{\citenamefont {Markovi\'c}\ \emph {et~al.}(2019)\citenamefont
  {Markovi\'c}, \citenamefont {Leroux}, \citenamefont {Riou}, \citenamefont
  {Araujo}, \citenamefont {Torrejon}, \citenamefont {Querlioz}, \citenamefont
  {Fukushima}, \citenamefont {Yuasa}, \citenamefont {Trastoy}, \citenamefont
  {Bortolotti},\ and\ \citenamefont {Grollier}}]{markovic19}%
  \BibitemOpen
  \bibfield  {author} {\bibinfo {author} {\bibfnamefont {D.}~\bibnamefont
  {Markovi\'c}}, \bibinfo {author} {\bibfnamefont {N.}~\bibnamefont {Leroux}},
  \bibinfo {author} {\bibfnamefont {M.}~\bibnamefont {Riou}}, \bibinfo {author}
  {\bibfnamefont {F.~A.}\ \bibnamefont {Araujo}}, \bibinfo {author}
  {\bibfnamefont {J.}~\bibnamefont {Torrejon}}, \bibinfo {author}
  {\bibfnamefont {D.}~\bibnamefont {Querlioz}}, \bibinfo {author}
  {\bibfnamefont {A.}~\bibnamefont {Fukushima}}, \bibinfo {author}
  {\bibfnamefont {S.}~\bibnamefont {Yuasa}}, \bibinfo {author} {\bibfnamefont
  {J.}~\bibnamefont {Trastoy}}, \bibinfo {author} {\bibfnamefont
  {P.}~\bibnamefont {Bortolotti}},\ and\ \bibinfo {author} {\bibfnamefont
  {J.}~\bibnamefont {Grollier}},\ }\bibfield  {title} {\bibinfo {title}
  {Reservoir computing with the frequency, phase, and amplitude of spin-torque
  nano-oscillators},\ }\href@noop {} {\bibfield  {journal} {\bibinfo  {journal}
  {Appl. Phys. Lett.}\ }\textbf {\bibinfo {volume} {114}},\ \bibinfo {pages}
  {012409} (\bibinfo {year} {2019})}\BibitemShut {NoStop}%
\bibitem [{\citenamefont {Tsunegi}\ \emph {et~al.}(2019)\citenamefont
  {Tsunegi}, \citenamefont {Taniguchi}, \citenamefont {Nakajima}, \citenamefont
  {Miwa}, \citenamefont {Yakushiji}, \citenamefont {Fukushima}, \citenamefont
  {Yuasa},\ and\ \citenamefont {Kubota}}]{tsunegi19}%
  \BibitemOpen
  \bibfield  {author} {\bibinfo {author} {\bibfnamefont {S.}~\bibnamefont
  {Tsunegi}}, \bibinfo {author} {\bibfnamefont {T.}~\bibnamefont {Taniguchi}},
  \bibinfo {author} {\bibfnamefont {K.}~\bibnamefont {Nakajima}}, \bibinfo
  {author} {\bibfnamefont {S.}~\bibnamefont {Miwa}}, \bibinfo {author}
  {\bibfnamefont {K.}~\bibnamefont {Yakushiji}}, \bibinfo {author}
  {\bibfnamefont {A.}~\bibnamefont {Fukushima}}, \bibinfo {author}
  {\bibfnamefont {S.}~\bibnamefont {Yuasa}},\ and\ \bibinfo {author}
  {\bibfnamefont {H.}~\bibnamefont {Kubota}},\ }\bibfield  {title} {\bibinfo
  {title} {Physical reservoir computing based on spin torque oscillator with
  forced synchronization},\ }\href@noop {} {\bibfield  {journal} {\bibinfo
  {journal} {Appl. Phys. Lett.}\ }\textbf {\bibinfo {volume} {114}},\ \bibinfo
  {pages} {164101} (\bibinfo {year} {2019})}\BibitemShut {NoStop}%
\bibitem [{\citenamefont {Riou}\ \emph {et~al.}(2019)\citenamefont {Riou},
  \citenamefont {Torrejon}, \citenamefont {Garitaine}, \citenamefont {Araujo},
  \citenamefont {Bortolotti}, \citenamefont {Cros}, \citenamefont {Tsunegi},
  \citenamefont {Yakushiji}, \citenamefont {Fukushima}, \citenamefont {Kubota},
  \citenamefont {Yuasa}, \citenamefont {Querlioz}, \citenamefont {Stiles},\
  and\ \citenamefont {Grollier}}]{riou19}%
  \BibitemOpen
  \bibfield  {author} {\bibinfo {author} {\bibfnamefont {M.}~\bibnamefont
  {Riou}}, \bibinfo {author} {\bibfnamefont {J.}~\bibnamefont {Torrejon}},
  \bibinfo {author} {\bibfnamefont {B.}~\bibnamefont {Garitaine}}, \bibinfo
  {author} {\bibfnamefont {F.~A.}\ \bibnamefont {Araujo}}, \bibinfo {author}
  {\bibfnamefont {P.}~\bibnamefont {Bortolotti}}, \bibinfo {author}
  {\bibfnamefont {V.}~\bibnamefont {Cros}}, \bibinfo {author} {\bibfnamefont
  {S.}~\bibnamefont {Tsunegi}}, \bibinfo {author} {\bibfnamefont
  {K.}~\bibnamefont {Yakushiji}}, \bibinfo {author} {\bibfnamefont
  {A.}~\bibnamefont {Fukushima}}, \bibinfo {author} {\bibfnamefont
  {H.}~\bibnamefont {Kubota}}, \bibinfo {author} {\bibfnamefont
  {S.}~\bibnamefont {Yuasa}}, \bibinfo {author} {\bibfnamefont
  {D.}~\bibnamefont {Querlioz}}, \bibinfo {author} {\bibfnamefont {M.~D.}\
  \bibnamefont {Stiles}},\ and\ \bibinfo {author} {\bibfnamefont
  {J.}~\bibnamefont {Grollier}},\ }\bibfield  {title} {\bibinfo {title}
  {Temporal {Patter} {Recognition} with {Delayed}-{Feedback} {Spin}-{Torque}
  {Nano}-{Oscillators}},\ }\href@noop {} {\bibfield  {journal} {\bibinfo
  {journal} {Phys. Rev. Applied}\ }\textbf {\bibinfo {volume} {12}},\ \bibinfo
  {pages} {024049} (\bibinfo {year} {2019})}\BibitemShut {NoStop}%
\bibitem [{\citenamefont {Yamaguchi}\ \emph {et~al.}(2020)\citenamefont
  {Yamaguchi}, \citenamefont {Akashi}, \citenamefont {Tsunegi}, \citenamefont
  {Kubota}, \citenamefont {Nakajima},\ and\ \citenamefont
  {Taniguchi}}]{yamaguchi20}%
  \BibitemOpen
  \bibfield  {author} {\bibinfo {author} {\bibfnamefont {T.}~\bibnamefont
  {Yamaguchi}}, \bibinfo {author} {\bibfnamefont {N.}~\bibnamefont {Akashi}},
  \bibinfo {author} {\bibfnamefont {S.}~\bibnamefont {Tsunegi}}, \bibinfo
  {author} {\bibfnamefont {H.}~\bibnamefont {Kubota}}, \bibinfo {author}
  {\bibfnamefont {K.}~\bibnamefont {Nakajima}},\ and\ \bibinfo {author}
  {\bibfnamefont {T.}~\bibnamefont {Taniguchi}},\ }\bibfield  {title} {\bibinfo
  {title} {Periodic structure of memory function in spintronics reservoir with
  feedback current},\ }\href@noop {} {\bibfield  {journal} {\bibinfo  {journal}
  {Phys. Rev. Research}\ }\textbf {\bibinfo {volume} {2}},\ \bibinfo {pages}
  {023389} (\bibinfo {year} {2020})}\BibitemShut {NoStop}%
\bibitem [{\citenamefont {Slonczewski}(1996)}]{slonczewski96}%
  \BibitemOpen
  \bibfield  {author} {\bibinfo {author} {\bibfnamefont {J.~C.}\ \bibnamefont
  {Slonczewski}},\ }\bibfield  {title} {\bibinfo {title} {Current-driven
  excitation of magnetic multilayers},\ }\href@noop {} {\bibfield  {journal}
  {\bibinfo  {journal} {J. Magn. Magn. Mater.}\ }\textbf {\bibinfo {volume}
  {159}},\ \bibinfo {pages} {L1} (\bibinfo {year} {1996})}\BibitemShut
  {NoStop}%
\bibitem [{\citenamefont {Berger}(1996)}]{berger96}%
  \BibitemOpen
  \bibfield  {author} {\bibinfo {author} {\bibfnamefont {L.}~\bibnamefont
  {Berger}},\ }\bibfield  {title} {\bibinfo {title} {Emission of spin waves by
  a magnetic multilayer traversed by a current},\ }\href@noop {} {\bibfield
  {journal} {\bibinfo  {journal} {Phys. Rev. B}\ }\textbf {\bibinfo {volume}
  {54}},\ \bibinfo {pages} {9353} (\bibinfo {year} {1996})}\BibitemShut
  {NoStop}%
\bibitem [{\citenamefont {Katine}\ \emph {et~al.}(2000)\citenamefont {Katine},
  \citenamefont {Albert}, \citenamefont {Buhrman}, \citenamefont {Myers},\ and\
  \citenamefont {Ralph}}]{katine00}%
  \BibitemOpen
  \bibfield  {author} {\bibinfo {author} {\bibfnamefont {J.~A.}\ \bibnamefont
  {Katine}}, \bibinfo {author} {\bibfnamefont {F.~J.}\ \bibnamefont {Albert}},
  \bibinfo {author} {\bibfnamefont {R.~A.}\ \bibnamefont {Buhrman}}, \bibinfo
  {author} {\bibfnamefont {E.~B.}\ \bibnamefont {Myers}},\ and\ \bibinfo
  {author} {\bibfnamefont {D.~C.}\ \bibnamefont {Ralph}},\ }\bibfield  {title}
  {\bibinfo {title} {Current-{Driven} {Magnetization} {Reversal} and
  {Spin}-{Wave} {Excitations} in {Co}/ {Cu}/{Co} {Pillars}},\ }\href@noop {}
  {\bibfield  {journal} {\bibinfo  {journal} {Phys. Rev. Lett.}\ }\textbf
  {\bibinfo {volume} {84}},\ \bibinfo {pages} {3149} (\bibinfo {year}
  {2000})}\BibitemShut {NoStop}%
\bibitem [{\citenamefont {Kiselev}\ \emph {et~al.}(2003)\citenamefont
  {Kiselev}, \citenamefont {Sankey}, \citenamefont {Krivorotov}, \citenamefont
  {Emley}, \citenamefont {Schoelkopf}, \citenamefont {Buhrman},\ and\
  \citenamefont {Ralph}}]{kiselev03}%
  \BibitemOpen
  \bibfield  {author} {\bibinfo {author} {\bibfnamefont {S.~I.}\ \bibnamefont
  {Kiselev}}, \bibinfo {author} {\bibfnamefont {J.~C.}\ \bibnamefont {Sankey}},
  \bibinfo {author} {\bibfnamefont {I.~N.}\ \bibnamefont {Krivorotov}},
  \bibinfo {author} {\bibfnamefont {N.~C.}\ \bibnamefont {Emley}}, \bibinfo
  {author} {\bibfnamefont {R.~J.}\ \bibnamefont {Schoelkopf}}, \bibinfo
  {author} {\bibfnamefont {R.~A.}\ \bibnamefont {Buhrman}},\ and\ \bibinfo
  {author} {\bibfnamefont {D.~C.}\ \bibnamefont {Ralph}},\ }\bibfield  {title}
  {\bibinfo {title} {Microwave oscillations of a nanomagnet driven by a
  spin-polarized current},\ }\href@noop {} {\bibfield  {journal} {\bibinfo
  {journal} {Nature}\ }\textbf {\bibinfo {volume} {425}},\ \bibinfo {pages}
  {380} (\bibinfo {year} {2003})}\BibitemShut {NoStop}%
\bibitem [{\citenamefont {Rippard}\ \emph {et~al.}(2004)\citenamefont
  {Rippard}, \citenamefont {Pufall}, \citenamefont {Kaka}, \citenamefont
  {Russek},\ and\ \citenamefont {Silva}}]{rippard04}%
  \BibitemOpen
  \bibfield  {author} {\bibinfo {author} {\bibfnamefont {W.~H.}\ \bibnamefont
  {Rippard}}, \bibinfo {author} {\bibfnamefont {M.~R.}\ \bibnamefont {Pufall}},
  \bibinfo {author} {\bibfnamefont {S.}~\bibnamefont {Kaka}}, \bibinfo {author}
  {\bibfnamefont {S.~E.}\ \bibnamefont {Russek}},\ and\ \bibinfo {author}
  {\bibfnamefont {T.~J.}\ \bibnamefont {Silva}},\ }\bibfield  {title} {\bibinfo
  {title} {Direct-{C}urrent {I}nduced {D}ynamics in
  {C}o${}_{90}${F}e${}_{10}$/{N}i${}_{80}${F}e${}_{20}$ {P}oint {C}ontacts},\
  }\href@noop {} {\bibfield  {journal} {\bibinfo  {journal} {Phys. Rev. Lett.}\
  }\textbf {\bibinfo {volume} {92}},\ \bibinfo {pages} {027201} (\bibinfo
  {year} {2004})}\BibitemShut {NoStop}%
\bibitem [{\citenamefont {Kubota}\ \emph {et~al.}(2013)\citenamefont {Kubota},
  \citenamefont {Yakushiji}, \citenamefont {Fukushima}, \citenamefont {Tamaru},
  \citenamefont {Konoto}, \citenamefont {Nozaki}, \citenamefont {Ishibashi},
  \citenamefont {Saruya}, \citenamefont {Yuasa}, \citenamefont {Taniguchi},
  \citenamefont {Arai},\ and\ \citenamefont {Imamura}}]{kubota13}%
  \BibitemOpen
  \bibfield  {author} {\bibinfo {author} {\bibfnamefont {H.}~\bibnamefont
  {Kubota}}, \bibinfo {author} {\bibfnamefont {K.}~\bibnamefont {Yakushiji}},
  \bibinfo {author} {\bibfnamefont {A.}~\bibnamefont {Fukushima}}, \bibinfo
  {author} {\bibfnamefont {S.}~\bibnamefont {Tamaru}}, \bibinfo {author}
  {\bibfnamefont {M.}~\bibnamefont {Konoto}}, \bibinfo {author} {\bibfnamefont
  {T.}~\bibnamefont {Nozaki}}, \bibinfo {author} {\bibfnamefont
  {S.}~\bibnamefont {Ishibashi}}, \bibinfo {author} {\bibfnamefont
  {T.}~\bibnamefont {Saruya}}, \bibinfo {author} {\bibfnamefont
  {S.}~\bibnamefont {Yuasa}}, \bibinfo {author} {\bibfnamefont
  {T.}~\bibnamefont {Taniguchi}}, \bibinfo {author} {\bibfnamefont
  {H.}~\bibnamefont {Arai}},\ and\ \bibinfo {author} {\bibfnamefont
  {H.}~\bibnamefont {Imamura}},\ }\bibfield  {title} {\bibinfo {title}
  {Spin-{T}orque {O}scillator {B}ased on {M}agnetic {T}unnel {J}unction with a
  {P}erpendicularly {M}agnetized {F}ree {L}ayer and {I}n-plane {M}agnetized
  {P}olarizer},\ }\href@noop {} {\bibfield  {journal} {\bibinfo  {journal}
  {Appl. Phys. Express}\ }\textbf {\bibinfo {volume} {6}},\ \bibinfo {pages}
  {103003} (\bibinfo {year} {2013})}\BibitemShut {NoStop}%
\bibitem [{\citenamefont {Thiele}(1973)}]{thiele73}%
  \BibitemOpen
  \bibfield  {author} {\bibinfo {author} {\bibfnamefont {A.~A.}\ \bibnamefont
  {Thiele}},\ }\bibfield  {title} {\bibinfo {title} {Steady-{S}tate {M}otion of
  {M}agnetic {D}omains},\ }\href@noop {} {\bibfield  {journal} {\bibinfo
  {journal} {Phys. Rev. Lett.}\ }\textbf {\bibinfo {volume} {30}},\ \bibinfo
  {pages} {230} (\bibinfo {year} {1973})}\BibitemShut {NoStop}%
\bibitem [{\citenamefont {Guslienko}\ \emph {et~al.}(2006)\citenamefont
  {Guslienko}, \citenamefont {Han}, \citenamefont {Keavney}, \citenamefont
  {Divan},\ and\ \citenamefont {Bader}}]{guslienko06PRL}%
  \BibitemOpen
  \bibfield  {author} {\bibinfo {author} {\bibfnamefont {K.~Y.}\ \bibnamefont
  {Guslienko}}, \bibinfo {author} {\bibfnamefont {X.~F.}\ \bibnamefont {Han}},
  \bibinfo {author} {\bibfnamefont {D.~J.}\ \bibnamefont {Keavney}}, \bibinfo
  {author} {\bibfnamefont {R.}~\bibnamefont {Divan}},\ and\ \bibinfo {author}
  {\bibfnamefont {S.~D.}\ \bibnamefont {Bader}},\ }\bibfield  {title} {\bibinfo
  {title} {Magnetic {Vortex} {Core} {Dynamics} in {Cylindrical} {Ferromagnetic}
  {Dots}},\ }\href@noop {} {\bibfield  {journal} {\bibinfo  {journal} {Phys.
  Rev. Lett.}\ }\textbf {\bibinfo {volume} {96}},\ \bibinfo {pages} {067205}
  (\bibinfo {year} {2006})}\BibitemShut {NoStop}%
\bibitem [{\citenamefont {Guslienko}(2006)}]{guslienko06}%
  \BibitemOpen
  \bibfield  {author} {\bibinfo {author} {\bibfnamefont {K.~Y.}\ \bibnamefont
  {Guslienko}},\ }\bibfield  {title} {\bibinfo {title} {Low-frequency vortex
  dynamic susceptibility and relaxation in mesoscopic ferromagnetic dots},\
  }\href@noop {} {\bibfield  {journal} {\bibinfo  {journal} {Appl. Phys.
  Lett.}\ }\textbf {\bibinfo {volume} {89}},\ \bibinfo {pages} {022510}
  (\bibinfo {year} {2006})}\BibitemShut {NoStop}%
\bibitem [{\citenamefont {Khvalkovskiy}\ \emph {et~al.}(2009)\citenamefont
  {Khvalkovskiy}, \citenamefont {Grollier}, \citenamefont {Dussaux},
  \citenamefont {Zvezdin},\ and\ \citenamefont {Cros}}]{khvalkovskiy09}%
  \BibitemOpen
  \bibfield  {author} {\bibinfo {author} {\bibfnamefont {A.~V.}\ \bibnamefont
  {Khvalkovskiy}}, \bibinfo {author} {\bibfnamefont {J.}~\bibnamefont
  {Grollier}}, \bibinfo {author} {\bibfnamefont {A.}~\bibnamefont {Dussaux}},
  \bibinfo {author} {\bibfnamefont {K.~A.}\ \bibnamefont {Zvezdin}},\ and\
  \bibinfo {author} {\bibfnamefont {V.}~\bibnamefont {Cros}},\ }\bibfield
  {title} {\bibinfo {title} {Vortex oscillations induced by spin-polarized
  current in a magnetic nanopillar: {A}nalytical versus micromagnetic
  calculations},\ }\href@noop {} {\bibfield  {journal} {\bibinfo  {journal}
  {Phys. Rev. B}\ }\textbf {\bibinfo {volume} {80}},\ \bibinfo {pages}
  {140401(R)} (\bibinfo {year} {2009})}\BibitemShut {NoStop}%
\bibitem [{\citenamefont {Guslienko}\ \emph {et~al.}(2011)\citenamefont
  {Guslienko}, \citenamefont {Aranda},\ and\ \citenamefont
  {Gonzalez}}]{guslienko11}%
  \BibitemOpen
  \bibfield  {author} {\bibinfo {author} {\bibfnamefont {K.~Y.}\ \bibnamefont
  {Guslienko}}, \bibinfo {author} {\bibfnamefont {G.~R.}\ \bibnamefont
  {Aranda}},\ and\ \bibinfo {author} {\bibfnamefont {J.}~\bibnamefont
  {Gonzalez}},\ }\bibfield  {title} {\bibinfo {title} {Spin torque and critical
  currents for magnetic vortex nano-oscillator in nanopillars},\ }\href@noop {}
  {\bibfield  {journal} {\bibinfo  {journal} {J. Phys. Conf. Ser.}\ }\textbf
  {\bibinfo {volume} {292}},\ \bibinfo {pages} {012006} (\bibinfo {year}
  {2011})}\BibitemShut {NoStop}%
\bibitem [{\citenamefont {Dussaux}\ \emph {et~al.}(2012)\citenamefont
  {Dussaux}, \citenamefont {Khvalkovskiy}, \citenamefont {Bortolotti},
  \citenamefont {Grollier}, \citenamefont {Cros},\ and\ \citenamefont
  {Fert}}]{dussaux12}%
  \BibitemOpen
  \bibfield  {author} {\bibinfo {author} {\bibfnamefont {A.}~\bibnamefont
  {Dussaux}}, \bibinfo {author} {\bibfnamefont {A.~V.}\ \bibnamefont
  {Khvalkovskiy}}, \bibinfo {author} {\bibfnamefont {P.}~\bibnamefont
  {Bortolotti}}, \bibinfo {author} {\bibfnamefont {J.}~\bibnamefont
  {Grollier}}, \bibinfo {author} {\bibfnamefont {V.}~\bibnamefont {Cros}},\
  and\ \bibinfo {author} {\bibfnamefont {A.}~\bibnamefont {Fert}},\ }\bibfield
  {title} {\bibinfo {title} {Field dependence of spin-transfer-induced vortex
  dynamics in the nonlinear regime},\ }\href@noop {} {\bibfield  {journal}
  {\bibinfo  {journal} {Phys. Rev. B}\ }\textbf {\bibinfo {volume} {86}},\
  \bibinfo {pages} {014402} (\bibinfo {year} {2012})}\BibitemShut {NoStop}%
\bibitem [{\citenamefont {Grimaldi}\ \emph {et~al.}(2014)\citenamefont
  {Grimaldi}, \citenamefont {Dussaux}, \citenamefont {Bortolotti},
  \citenamefont {Grollier}, \citenamefont {Pillet}, \citenamefont {Fukushima},
  \citenamefont {Kubota}, \citenamefont {Yakushiji}, \citenamefont {Yuasa},\
  and\ \citenamefont {Cros}}]{grimaldi14}%
  \BibitemOpen
  \bibfield  {author} {\bibinfo {author} {\bibfnamefont {E.}~\bibnamefont
  {Grimaldi}}, \bibinfo {author} {\bibfnamefont {A.}~\bibnamefont {Dussaux}},
  \bibinfo {author} {\bibfnamefont {P.}~\bibnamefont {Bortolotti}}, \bibinfo
  {author} {\bibfnamefont {J.}~\bibnamefont {Grollier}}, \bibinfo {author}
  {\bibfnamefont {G.}~\bibnamefont {Pillet}}, \bibinfo {author} {\bibfnamefont
  {A.}~\bibnamefont {Fukushima}}, \bibinfo {author} {\bibfnamefont
  {H.}~\bibnamefont {Kubota}}, \bibinfo {author} {\bibfnamefont
  {K.}~\bibnamefont {Yakushiji}}, \bibinfo {author} {\bibfnamefont
  {S.}~\bibnamefont {Yuasa}},\ and\ \bibinfo {author} {\bibfnamefont
  {V.}~\bibnamefont {Cros}},\ }\bibfield  {title} {\bibinfo {title} {Response
  to noise of a vortex based spin transfer nano-oscillator},\ }\href@noop {}
  {\bibfield  {journal} {\bibinfo  {journal} {Phys. Rev. B}\ }\textbf {\bibinfo
  {volume} {89}},\ \bibinfo {pages} {104404} (\bibinfo {year}
  {2014})}\BibitemShut {NoStop}%
\bibitem [{\citenamefont {Tsunegi}\ \emph {et~al.}(2014)\citenamefont
  {Tsunegi}, \citenamefont {Kubota}, \citenamefont {Yakushiji}, \citenamefont
  {Konoto}, \citenamefont {Tamaru}, \citenamefont {Fukushima}, \citenamefont
  {Arai}, \citenamefont {Imamura}, \citenamefont {Grimaldi}, \citenamefont
  {Lebrun}, \citenamefont {Grollier}, \citenamefont {Cros},\ and\ \citenamefont
  {Yuasa}}]{tsunegi14}%
  \BibitemOpen
  \bibfield  {author} {\bibinfo {author} {\bibfnamefont {S.}~\bibnamefont
  {Tsunegi}}, \bibinfo {author} {\bibfnamefont {H.}~\bibnamefont {Kubota}},
  \bibinfo {author} {\bibfnamefont {K.}~\bibnamefont {Yakushiji}}, \bibinfo
  {author} {\bibfnamefont {M.}~\bibnamefont {Konoto}}, \bibinfo {author}
  {\bibfnamefont {S.}~\bibnamefont {Tamaru}}, \bibinfo {author} {\bibfnamefont
  {A.}~\bibnamefont {Fukushima}}, \bibinfo {author} {\bibfnamefont
  {H.}~\bibnamefont {Arai}}, \bibinfo {author} {\bibfnamefont {H.}~\bibnamefont
  {Imamura}}, \bibinfo {author} {\bibfnamefont {E.}~\bibnamefont {Grimaldi}},
  \bibinfo {author} {\bibfnamefont {R.}~\bibnamefont {Lebrun}}, \bibinfo
  {author} {\bibfnamefont {J.}~\bibnamefont {Grollier}}, \bibinfo {author}
  {\bibfnamefont {V.}~\bibnamefont {Cros}},\ and\ \bibinfo {author}
  {\bibfnamefont {S.}~\bibnamefont {Yuasa}},\ }\bibfield  {title} {\bibinfo
  {title} {High emission power and {Q} factor in spin torque vortex oscillator
  consisting of {Fe}{B} free layer},\ }\href@noop {} {\bibfield  {journal}
  {\bibinfo  {journal} {Appl. Phys. Express}\ }\textbf {\bibinfo {volume}
  {7}},\ \bibinfo {pages} {063009} (\bibinfo {year} {2014})}\BibitemShut
  {NoStop}%
\bibitem [{\citenamefont {Tsunegi}\ \emph {et~al.}(2016)\citenamefont
  {Tsunegi}, \citenamefont {Yakushiji}, \citenamefont {Fukushima},
  \citenamefont {Yuasa},\ and\ \citenamefont {Kubota}}]{tsunegi16}%
  \BibitemOpen
  \bibfield  {author} {\bibinfo {author} {\bibfnamefont {S.}~\bibnamefont
  {Tsunegi}}, \bibinfo {author} {\bibfnamefont {K.}~\bibnamefont {Yakushiji}},
  \bibinfo {author} {\bibfnamefont {A.}~\bibnamefont {Fukushima}}, \bibinfo
  {author} {\bibfnamefont {S.}~\bibnamefont {Yuasa}},\ and\ \bibinfo {author}
  {\bibfnamefont {H.}~\bibnamefont {Kubota}},\ }\bibfield  {title} {\bibinfo
  {title} {Microwave emission power exceeding 10 $\mu${W} in spin torque vortex
  oscillator},\ }\href@noop {} {\bibfield  {journal} {\bibinfo  {journal}
  {Appl. Phys. Lett.}\ }\textbf {\bibinfo {volume} {109}},\ \bibinfo {pages}
  {252402} (\bibinfo {year} {2016})}\BibitemShut {NoStop}%
\bibitem [{\citenamefont {Strogatz}(2001)}]{strogatz01}%
  \BibitemOpen
  \bibfield  {author} {\bibinfo {author} {\bibfnamefont {S.~H.}\ \bibnamefont
  {Strogatz}},\ }\href@noop {} {\emph {\bibinfo {title} {Nonlinear Dynamics and
  Chaos: With Applications to Physics, Biology, Chemistry, and Engineering}}},\
  \bibinfo {edition} {1st}\ ed.\ (\bibinfo  {publisher} {Westview Press},\
  \bibinfo {year} {2001})\BibitemShut {NoStop}%
\bibitem [{\citenamefont {Pikovsky}\ \emph {et~al.}(2003)\citenamefont
  {Pikovsky}, \citenamefont {Rosenblum},\ and\ \citenamefont
  {Kurths}}]{pikovsky03}%
  \BibitemOpen
  \bibfield  {author} {\bibinfo {author} {\bibfnamefont {A.}~\bibnamefont
  {Pikovsky}}, \bibinfo {author} {\bibfnamefont {M.}~\bibnamefont
  {Rosenblum}},\ and\ \bibinfo {author} {\bibfnamefont {J.}~\bibnamefont
  {Kurths}},\ }\href@noop {} {\emph {\bibinfo {title} {Synchronization: A
  universal concept in nonlinear sciences}}},\ \bibinfo {edition} {1st}\ ed.\
  (\bibinfo  {publisher} {Cambridge University Press},\ \bibinfo {year}
  {2003})\BibitemShut {NoStop}%
\end{thebibliography}


\end{document}